\documentclass[12pt]{article}
\usepackage{epsf,epsfig}

\newcommand{\autor}[1] {\begin{center}{\bf \lineskip .3cm #1} \end{center}}
\newcommand{\address}[1] {\begin{center}  {\normalsize \bf \it #1 }\end{center}}
\def\simgt{\rlap{\lower 3.5 pt \hbox{$\mathchar \sim$}} \raise 1pt \hbox {$>$}}
\def\simlt{\rlap{\lower 3.5 pt \hbox{$\mathchar \sim$}} \raise 1pt \hbox {$<$}}

\def\3half{\textstyle\frac32}
\begin{document}
\begin{titlepage}
\vspace{1.0cm}
\begin{center}
\large\bf\boldmath Form Factor for $B{\to}D l\tilde{\nu}$ in The
Light-Cone Sum Rules With Chiral Current Correlator\unboldmath
\end{center}
\vspace{0.8cm} \autor{{Fen Zuo,$^{1,~}$\footnote{Email:
zuof@mail.ihep.ac.cn}}~Zuo-Hong Li $^{2,~3}$\footnote{Email:
lizh@ytu.edu.cn} and Tao Huang $^{1,~3}$\footnote{Email:
huangtao@mail.ihep.ac.cn}} \vspace{0.7cm}
\address{$^1$ Institute of High Energy Physics, P.O.Box 918 ,
Beijing 100049, P.R.China\,\footnote{Mailing address}}
\address{$^2$ Department of Physics, Yantai University,
Yantai 264005,China}
\address{$^3$ CCAST (World Laboratory), P.O.Box 8730, Beijing 100080,
China} \vspace{1.0cm}
\begin{abstract}
Within the framework of  QCD light-cone sum rules (LCSR), we
calculate the form factor for $B{\to}D l\tilde{\nu}$ transitions
with chiral current correlator. The resulting form factor depends on
the distribution amplitude (DA) of the $D$-meson. We try to use
three kinds of DA models of the $D$-meson. In the velocity transfer
region $1.14 < y < 1.59$, which renders the Operator Product
Expansion (OPE) near light-cone $x^2=0$ go effectively, the yielding
behavior of form factor is in agreement with that extracted from the
data on $ B\to D l\tilde{\nu}$, within the error. In the large
recoil region $1.35 < y < 1.59$, the results are observed consistent
with those of perturbative QCD (pQCD). The presented calculation can
play a bridge role connecting those from the lattice QCD, heavy
quark symmetry and pQCD to have an all-around understanding of
$B{\to}D l\tilde{\nu}$ transitions.
\end{abstract}
\end{titlepage}

\section{Introduction}

~~~~Calculation of the form factors for semileptonic transitions of
$B$ mesons has been being a subject discussed intensely. Recently,
it has been shown that the $B\to\pi$ transition form factor can be
consistently analyzed by using the different approaches in the
different $q^2$ regions \cite{Bpi1, cLCSR2, Bpi2, Bpi3}. The
perturbative QCD (pQCD) can be applied to the $B\to\pi$ form factor
in the large recoil (small $q^2$) region and it is reliable when the
involved energy scale is large enough \cite{Bpi1}. The QCD
light-cone sum rules (LCSR) can involve both the hard and soft
contributions to the $B\to\pi$ form factor below
$q^2\simeq18~\mbox{GeV}^2$ \cite{cLCSR2}. The lattice QCD
simulations of the $B\to\pi$ transition form factor \cite{Bpi2} are
available only for the soft region $q^2>15~\mbox{GeV}^2$, because of
the restriction to the $\pi$ energy smaller than the inverse lattice
spacing. Thus the results from these three approaches might be
complementary to each other. In Ref.\cite{Bpi3} we recalculate the
$B\to\pi$ form factor in the pQCD approach, with the transverse
momentum dependence included for both the hard scattering part and
the nonperturbative wave functions(of $\pi$ and $B$) to get a more
reliable pQCD result. By combining the results from these three
methods we obtain a full understanding of the $B\to\pi$ transition
form factor in its physical region
$0\leq{q^2}\leq(M_B-M_\pi)^2\simeq25~\mbox{GeV}^2$.

Up to now, in comparison with heavy-to-light cases the calculations
on heavy-to-heavy transitions can be done only for a certain
specific kinematical range, although there have been a lot of
discussions in the literature. In the BSW model \cite{BSW}, the
relevant form factors at zero momentum transfer are expressed as an
overlap of initial and final meson wave functions  for which they
take the solutions of the Bethe-Salpeter (BS) equation in a
relativistic harmonic oscillator potential. Then one extrapolates
the result at $q^2=0$ to the whole kinematical region assuming the
nearest pole dominance. With the discovering of the heavy quark
symmetry, the $B{\to}D$ form factor have been known better at zero
recoil. This is because of the fact that in the heavy quark limit
the resulting form factors~---~Isgur-Wise functions \cite{IS} at
zero recoil are rigorously normalized. Including the leading
symmetry breaking corrections, the deviation from this limit can be
estimated at an order of a few percent due to Luke's theorem and
therefore the value of the form factor at this point can be
determined within a higher accuracy \cite{Neubert1}. However, the
dependence of the form factor on the velocity transfer $y=v\cdot v'$
(with $v$ and $v'$ being the velocities of the $B$ and $D$ mesons
respectively) is difficult to get even in the leading order, in view
of the arbitrary function $\sigma(y)$ \cite{Neubert2} which is
introduced to simulate higher-resonances in the heavy quark
effective theory (HQET). The lattice QCD, despite a rigorous
nonperturbative approach, is just adequate to estimate the behavior
of the form factors near the zero recoil \cite{lattice}. Among the
other approaches are the QCD sum rules and pQCD. Ref.\cite{3P}
applies the traditional $3$-point sum rule to calculate the form
factor at zero momentum transfer. It is concluded in Ref.\cite{pQCD}
that pQCD approach is applicable in the large recoil region and can
give a consistent result with the experiment.

It is necessary that there is a reliable estimate of $B\to D$
transition in the whole kinematically accessible range
$0\leq{q^2}\leq(M_B-M_D)^2\simeq 11.6~\mbox{GeV}^2$, in order to
account for the data on $B{\to}D l\tilde{\nu}$. For this purpose, it
is practical, as shown in $B\to \pi$ case, to combine the result of
QCD LCSR with those from the lattice QCD, heavy quark symmetry and
pQCD. The LCSR approach \cite{LCSR}, where the non-perturbative
dynamics are effectively parameterized in so-called light-cone wave
functions, is regarded as an effective tool to deal with
heavy-to-light exclusive processes. Although the $B{\to}D$
transition in question is a heavy-to-heavy one, the $c$-quark is
much lighter compared to $b$-quark and so discussing it with LCSR is
plausible for the kinematical range where the OPE near light-cone
$x^2=0$ is valid. The other problem with our practical calculation
is that the higher twist DA's of D meson, which are important but
less studied, would enter into the sum rule result. However, an
effective approach \cite{cLCSR} has been presented to avoid the
pollution by some higher-twist DA's. This improved LSCR method uses
a certain chiral current correlator as the starting point so that
the relevant twist-$3$ wave functions make no contributions and the
reliability of calculation can be enhanced to a large extend. Its
applicability has been examined by a great deal of studies
\cite{cLCSR2,cLCSR1}. In this paper we would like to employ the
improved LCSR to discuss the form factor for the $B{\to}D$
transition and try to give a full understanding of QCD dynamics
involved in the $B\to D l\tilde{\nu}$.

This paper is organized as follows. In the following Section we
derive the LCSR for the form factor for $B\to D$. A discussion of
the DA models for the $D$-meson is given in section {I}{I}{I}.
Section {I}{V} is devoted to the numerical analysis and comparison
with other approaches. The last section is reserved for summary.

\section{Derivation of LCSR for The $B{\to}D$ Form Factor }
~~~The $B{\to}D$ weak form factors $f(q^2)$ and $\tilde{f}(q^2)$
are usually defined as:
\begin{equation}
{<}D(p)|\bar{c}\gamma_{\mu}b|B(p+q){>}=2f(q^2)p_\mu+\tilde{f}(q^2)q_\mu,\label{eq:def}
\end{equation}
with $q$ being the momentum transfer. On the other hand, when
applying the heavy quark symmetry to do discussion the following
definition is advisable,
\begin{equation}
{<}D(p)|\bar{c}\gamma_{\mu}b|B(p+q){>}=\sqrt{m_B~m_D}[h_+(y)(v+v')_\mu+h_-(y)(v-v')_\mu].
\end{equation}
If we neglect the masses of leptons in the decay final state of
$B{\to}Dl\tilde{\nu}_l $, only $f(q^2)$ is relevant and thus we can
confine us to the discussion on $f(q^2)$.  Obviously, the following
relation is observed between $f(q^2)$ and $h_{+ (-)}(y)$,
\begin{equation}
f(q^2)=\frac{m_B+m_D}{2\sqrt{m_B~m_D}}{\cal F}_{B\to
D}(y)\label{eq:f12}
\end{equation}
where ${\cal F}_{B\to D}(y)=h_+(y)-\frac{m_B-m_D}{m_B+m_D}h_-(y)$,
$q^2=m_B^2+m_D^2-2m_B~m_D y$.

Using the heavy quark symmetry, the value of form factor ${\cal
F}_{B\to D}(1)$ at zero recoil could be fixed better. Since in heavy
quark limit $h_+(1)=1$ and $h_-(1)=0$, the form factor ${\cal
F}_{B\to D}(1)$ should be close to $1$. A systematic investigation
gives ${\cal F}_{B\to D}(1)=0.98\pm 0.07$\cite{Neubert4}, with a
less model dependence. This result is also confirmed with lattice
calculations \cite{lattice}. PQCD analyses are also made in the
large recoil region $ y=1.35-1.59$, yielding a result consistent
with the data. The LCSR calculation can help to understand ${\cal
F}_{B\to D}(y)$ in the whole kinematical region in complementary to
the lattice QCD with the heavy quark symmetry and pQCD approaches.

To achieve a LCSR estimate of ${\cal F}_{B\to D}(y)$, we follow
\cite{cLCSR2} and use the following chiral current correlator
$\Pi_\mu(p,q)$:
\begin{eqnarray}
\Pi_\mu(p,q)&=&i\int{d^4xe^{ipx}{<}D(p)|T\{\bar{c}(x)\gamma_\mu(1+\gamma_5)b(x),
\bar{b}(0)i(1+\gamma_5)d(0)\}|0{>}}\nonumber\\
&=&\Pi(q^2,(p+q)^2)p_\mu+\tilde{\Pi}(q^2,(p+q)^2)q_\mu,\label{eq:cc}
\end{eqnarray}

In the first place, we discuss the hadronic representation for the
correlator. This can be done by inserting the complete intermediate
states with the same quantum numbers as the current operator
$\bar{b}i(1+\gamma_5)d$. Isolating the pole contribution due to the
lowest pseudoscalar $B$ meson, we have the hadronic representation
in the following:
\begin{eqnarray}
\Pi_{\mu}^H(p,q)&=&\Pi^H(q^2,(p+q)^2)p_\mu+\tilde{\Pi}^H(q^2,(p+q)^2)q_{\mu}\nonumber\\
&=&\frac{{<}D|\bar{c}\gamma_{\mu}b|B{>}{<}B|\bar{b}i\gamma_5d|0{>}}{m^2_B-(p+q)^2}\nonumber\\
&&+\sum_H\frac{{<}D|\bar{c}\gamma_{\mu}(1+\gamma_5)b|B^H{>}{<}B^H|\bar{b}i(1+\gamma_5)d|0{>}}{m^2_{B^H}-(p+q)^2}.
\end{eqnarray}
Note that the intermediate states $B^H$ contain not only the
pseudoscalar resonance of masses greater than $m_B$, but also the
scalar resonances with $J^P=0^+$, corresponding to the operator
$\bar{b}d$. With Eq.(\ref{eq:def}) and the definition of the decay
constant $f_B$ of the $B$-meson
\begin{equation}
{<}B|\bar{b}i\gamma_5d|0{>}=m_B^2f_B/m_b,
\end{equation}
and expressing the contributions of higher resonances and continuum
states in a form of dispersion integration, the invariant amplitudes
$\Pi^H$ and $\tilde\Pi^H$ read,
\begin{equation}
\Pi^H[q^2,(p+q)^2]=\frac{2f(q^2)m_B^2f_B}{m_b(m_B^2-(p+q)^2)}+
\int^\infty_{s_0}{\frac{\rho^H(s)}{s-(p+q)^2}ds}+\mbox{subtractions},
\end{equation}
and
\begin{equation}
\tilde{\Pi}^H[q^2,(p+q)^2]=\frac{\tilde{f}(q^2)m_B^2f_B}{m_b(m_B^2-(p+q)^2)}+
\int^\infty_{s_0}{\frac{\tilde{\rho}^H(s)}{s-(p+q)^2}ds}+\mbox{subtractions},
\end{equation}
where the threshold parameter $s_0$ should be set near the squared
mass of the lowest scalar $B$ meson, the spectral densities
$\rho^H(s)$ and $\tilde{\rho}^H(s)$ can be approximated by invoking
the quark-hadron duality ansatz
\begin{equation}
\rho^H(s)(\tilde\rho^H(s))=\rho^{QCD}(s)(\tilde\rho^{QCD}(s))\theta(s-s_0).
\end{equation}

On the other hand, we need to calculate the corrector in QCD theory
to obtain the desired sum rule result. In fact, there is an
effective kinematical region which makes OPE applicable:
$(p+q)^2-m_b^2{\ll}0$ for the $b\bar{d}$ channel and
$q^2{\le}(m_b-m_c)^2-2\Lambda_{QCD}(m_b-m_c)$ for the momentum
transfer.

For the present purpose, it is sufficient to consider the invariant
amplitude $\Pi(q^2,(p+q)^2)$ which contains the desired form factor.
The leading contribution is derived easily by contracting the
$b-$quark operators to a free propagator:
\begin{equation}
{<0}|Tb(x)\bar{b}(0)|{0>}=\int{\frac{d^4k}{{(2\pi)}^4}e^{-ikx}\frac{k\!\!\!/+m_b}{k^2-m_b^2}}.\label{eq:pro}
\end{equation}\
Substituting Eq.(\ref{eq:pro}) into Eq.(\ref{eq:cc}), we have the
two-particle contribution to the correlator,
\begin{equation}
\Pi_\mu^{(\bar{q}q)}=-2m_bi\int{\frac{d^4xd^4k}{(2\pi)^4}e^{i(q-k)x}\frac{1}{k^2-m_b^2}{<}D(p)|T\bar{c}(x)\gamma_{\mu}\gamma_5d(0)|{0>}}.\label{eq:qq}
\end{equation}
An important observation, as in Ref.\cite{cLCSR2}, is that only the
leading non-local matrix element
${<}D(p)|\bar{c}(x)\gamma_{\mu}\gamma_5d(0)|0{>}$ contributions to
the correlator, while the nonlocal matrix elements
${<}D(p)|\bar{c}(x)i\gamma_5d(0)|0{>}$ and
${<}D(p)|\bar{c}(x)\sigma_{\mu\nu}\gamma_5d(0)|0{>}$ whose leading
terms are of twist $3$, disappear from the sum rule. Proceeding to
Eq.(\ref{eq:qq}), we can expand the nonlocal matrix element
${<}D(p)|T\bar{c}(x)\gamma_{\mu}\gamma_5d(0)|0{>}$ as
\begin{equation}
{<}D(p)|T\bar{c}(x)\gamma_{\mu}\gamma_5d(0)|0{>}=-ip_{\mu}f_D\int_0^1{due^{iupx}\varphi_D(u)}+\mbox{higher
twist terms},\label{eq:da}
\end{equation}
where $\varphi_D(u)$ is the twist-2 DA of D meson with $u$ being the
longitudinal momentum fraction carried by the $c$-quark, those DA's
entering the higher-twist terms are of at least twist $4$. The use
of Eq.(\ref{eq:da}) yields
\begin{equation}
\Pi^{(\bar{q}q)}[q^2,(p+q)^2]=2f_Dm_b\int_0^1{du\frac{\varphi_D(u)}{m_b^2-(up+q)^2}}+\mbox{higher
twist terms}. \label{eq:qq1}
\end{equation}

Invoking a correction term due to the interaction of the b quark
with a background field gluon into (10), the three-particle
contribution $\Pi^{(\bar{q}qg)}_\mu$ is achievable. However, the
practical calculation shows that the corresponding matrix element
whose leading term is of twist $3$ also vanishes. Thus, If we work
to the twist-3 accuracy, only the leading twist DA $\varphi_D$ is
needed to yield a LCSR prediction.

Furthermore, we carry out the subtraction procedure of the continuum
spectrum, make the Borel transformations with respect to $(p+q)^2$
in the hadronic and the QCD expressions, and then equate them.
Finally, from Eq.(\ref{eq:f12}) follows the LCSR for ${\cal F}_{B\to
D}(y)$, which is applicable to the velocity transfer region $1.14< y
<1.59$,
\begin{eqnarray}
{\cal F}_{B\to
D}(y)&=&\frac{2m_b^2}{(m_B+m_D)m_B}\sqrt{\frac{m_D}{m_B}}\frac{f_D}{f_B}e^{m_B^2/M^2}\nonumber\\
&&\times\int_\Delta^1{\frac{du}{u}\exp{\left[-\frac{m_b^2-(1-u)(q^2-um_D^2)}{uM^2}\right]}
\varphi_D(u)},\label{eq:ff}
\end{eqnarray}
where
\begin{eqnarray}
\Delta=\frac{\sqrt{(s_0-q^2-m_D^2)^2+4m_D^2(m_b^2-q^2)}-(s_0-q^2-m_D^2)}{2m_D^2},
\end{eqnarray}
and $p^2=m_D^2$ has been used.

\section{$D$-meson Distribution Amplitude}

~~~Now let's do a brief discussion on an important nonperturbative
parameter appearing in the LCSR formula(\ref{eq:ff}), the leading
twist DA of $D$-meson, $\varphi_D(x)$.

$D$-meson is composed of the heavy quark $c$ and the light
anti-quark $\bar q$. The longitudinal momentum distribution should
be asymmetry and the peak of the distribution should be
approximately at $x\simeq0.7$. According to the definition in
Eq.(\ref{eq:da}), $\varphi_D(x)$ satisfies the normalization
condition
\begin{equation}
\int^1_0{dx \varphi_D(x)}=1\label{eq:constraint1},
\end{equation}
which is derived by the leptonic decay $D\to\mu\nu$.

In the pQCD calculations \cite{pQCD}, a simple model (we call model
I) is adopted as
\begin{equation}
\varphi_D^{(I)}(x)=6x(1-x)(1-C_d(1-2x))\label{eq:da1}
\end{equation}
which is based on the expansion of the Gegenbauer polynomials.
Eq.(\ref{eq:da1}) has a free parameter $C_d$ which ranges from $0$
to $1$. We will take $C_d=0.7$ as input.

On the other hand, it was suggested in \cite{wf} that the light-cone
wave function of the $D$-meson be taken as:
\begin{equation}
\psi_D(x,\mathbf{k}_\perp)=A_D\exp{\left[-b_D^2\left(\frac{\mathbf{k}_\perp^2+m_c^2}{x}
+\frac{\mathbf{k}_\perp^2+m_d^2}{1-x}\right)\right]}\label{eq:wf}
\end{equation}
which is derived from the Brosky-Huang-Lepage(BHL) prescription
\cite{BHL}. $\psi_D(x,\mathbf{k}_\perp)$ can be related to the DA by
the definition
\begin{equation}
\varphi_D(x)=\frac{2\sqrt{3}}{f_D}\int{\frac{d^2\mathbf{k}_\perp}{16\pi^3}
\psi_D(x,\mathbf{k}_\perp)}\label{eq:wf-da}.
\end{equation}
Substituting Eq.(\ref{eq:wf}) into Eq.(\ref{eq:wf-da}), we have a
model of the DA(model {I}{I})
\begin{equation}
\varphi_D^{({I}{I})}(x)=\frac{\sqrt{3}A_D}{8\pi^2~f_D~b_D^2}x(1-x)\exp{\left[-b_D^2\frac{x
m_d^2 +(1-x)m_c^2}{x(1-x)}\right]},\label{eq:da2}
\end{equation}
where the parameters $A_D$ and $b_D$ can be fixed by the
normalization(\ref{eq:constraint1}) and the probability of finding
the $|q\bar{q}>$ Fock state in the $D$- meson, $P_D$
\begin{equation}
P_D=\int^1_0{dx\int{\frac{d^2\mathbf{k}_\perp}{16\pi^3}|\psi_D(x,\mathbf{k}_{\perp})|^2}}\label{eq:constraint2}.
\end{equation}
As discussed in Ref.\cite{wf}, $P_D\approx0.8$ is a good
approximation for the $D$-meson (As we have checked, change of $P_D$
makes a numerical effect less than $2\%$). Then, taking
$P_D\approx{0.8}$, $f_D=240 \mbox{MeV}$, $m_c=1.3 \mbox{GeV}$ and
$m_d=0.35\mbox{GeV}$, we have $A_D=63.6\mbox{GeV}^{-1}$,
$b_D^2=0.292\mbox{GeV}^{-2}$.

Furthermore, as argued in Ref.\cite{pionwf}, a more complete form
of the light-cone wave function should include the Melosh rotation
effect in spin space:
\begin{equation}
\psi^f_D(x,\mathbf{k}_\perp)=
\chi_D(x,\mathbf{k}_\perp)\exp{\left[-b_D^2\left(\frac{\mathbf{k}_\perp^2+m_c^2}{x}
+\frac{\mathbf{k}_\perp^2+m_d^2}{1-x}\right)\right]}
\end{equation}
with the Melosh factor,
\begin{equation}
\chi_D(x,\mathbf{k}_\perp)=\frac{(1-x)m_c+xm_d}{\sqrt{\mathbf{k}^2_\perp+((1-x)m_c+xm_d)^2}}\label{eq:melosh}.
\end{equation}
It can be seen from Eq.(\ref{eq:melosh}) that
$\chi_D(x,\mathbf{k}_\perp)\to 1$ as $m_c\to\infty$, since there is
no spin interaction between the two quarks in the heavy-flavor
meson, ie., the spin of the heavy constituent decouples from the
gluon field, in the heavy quark limit \cite{IS}. However the
$c$-quark is not heavy enough to neglect the Melosh factor.

After integration over $\mathbf{k}_\perp$ the full form of $D$
meson DA can be achieved (model {I}{I}{I}):
\begin{equation}
\varphi_D^{({I}{I}{I})}(x)=\frac{A_D\sqrt{3x(1-x)}}{8\pi^{3/2}f_Db_D}y\left[1-Erf\left(\frac{b_Dy}{\sqrt{x(1-x)}}\right)\right]
\exp{\left[-b_D^2\frac{(xm_d^2+(1-x)m_c^2-y^2)}{x(1-x)}\right]}\label{eq:da3},
\end{equation}
where $y=xm_d+(1-x)m_c$ and the error function $Erf(x)$ is defined
as $Erf(x)=\frac{2}{\pi}\int^x_0{\exp({-t^2})dt}$. Using the same
constraints as in Eq.(\ref{eq:constraint1}) and
(\ref{eq:constraint2}), the parameters $A_D$ and $b_D$ are fixed as
$A_D=62.8\mbox{GeV}^{-1}$ and $b_D^2=0.265\mbox{GeV}^{-2}$.

\begin{figure}[p]
$$\epsfxsize=0.60\textwidth\epsffile{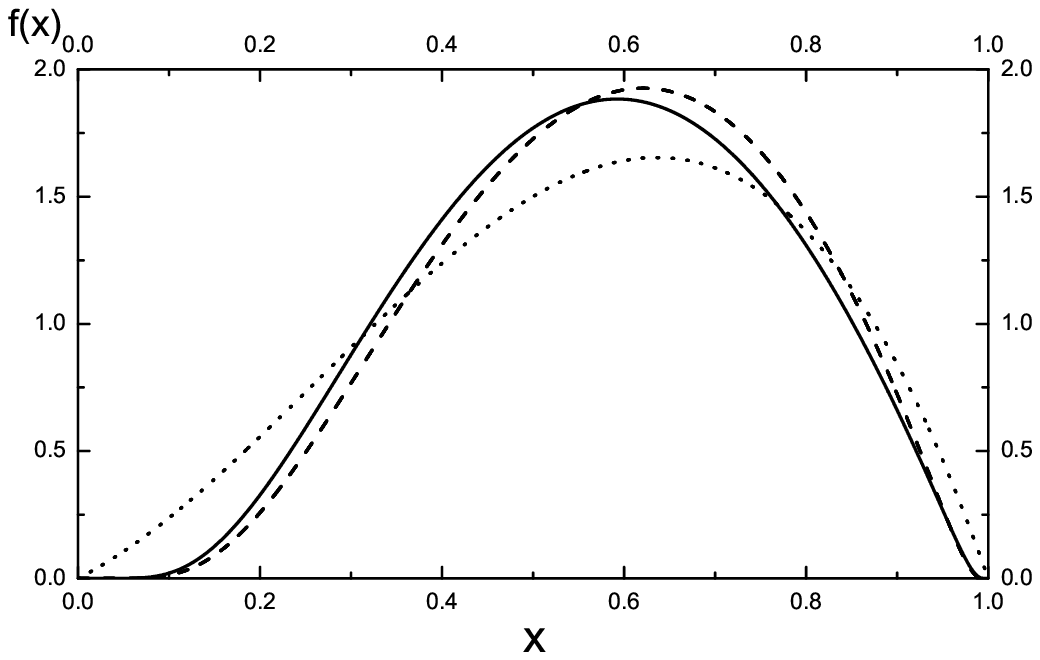}$$
\caption[]{Different kinds of $D$-meson DAs,solid and dashed curves
correspond to model {I}{I}{I} and {I}{I}, while the dotted line
expresses model I.}\label{fig:wf}
\end{figure}

In this paper we will employ the above three kinds of models to do
numerical calculation. All these DA's of the $D$-meson are plotted
in Fig.(\ref{fig:wf}) for a comparison. It is shown that they are
of similar shape and all of them exhibit a maximum at
$x\simeq0.6-0.7$ as expected.

\section{Numerical Result and Discussion}

~~~Apart from the DA of $D$-meson, the decay constant of $B$-meson
$f_B$ is among the important nonperturbative inputs. For
consistency, we use the following corrector
\begin{equation}
K(q^2)=i\int{d^4xe^{iqx}<0|\bar{q}(x)(1+\gamma_5)b(x),\bar{b}(0)(1-\gamma_5)q(0)|0>},
\end{equation}
to recalculate it in the two-point sum rules. The calculation should
be limited to leading order in QCD, since the QCD radiative
corrections to the sum rule for ${\cal F}_{B\to D}(y)$ are not taken
into account. The value of the threshold parameter $s'_0$ is
determined by a best fit requirement in the region $10
\mbox{GeV}^2{\leq}\bar{M}^2{\leq}20\mbox{GeV}^2$, where $\bar{M}^2$
is the corresponding Borel parameter. The same procedure is
performed for ${\cal F}_{B\to D}(y)$, resulting in different values
of the threshold parameter $s_0$. The result is listed in
Tab.\ref{tab:s0}. Choosing $s_0$ and $s'_0$ in this way, the
dependence of $f_B$ and ${\cal F}_{B\to D}(y)$ on the Borel
parameter is very weak and thus we can simply evaluate them at
$M^2=\bar{M}^2=15\mbox{GeV}^2$. The other input parameters are taken
as $m_B= 5.279\mbox{GeV}, m_D=1.869\mbox{GeV}$. As we have ignored
all the radiation corrections, we don't expect our values of $f_B$
to be good predictions of that quantity.

\begin{table}
\begin{center}
\begin{tabular}{|l|l|cc|c|}
\hline
      & $m_b$& $s'_0$ &$f_B$  & $s_0$  \\\hline
set 1 & 4.85 & 29.5  & 0.076 &  30.3  \\
set 2 & 4.80 & 29.8  & 0.090 &  30.8  \\
set 3 & 4.75 & 30.0  & 0.103 &  31.3  \\\hline
\end{tabular}
\end{center}
\caption[]{Parameter sets for $f_B$ and ${\cal F}_{B\to D}(y)$,
$s'_0$ and $s_0$ for $f_B$ and ${\cal F}_{B\to D}(y)$ respectively;
$m_b$ and $f_B$ are given in GeV, $s_0$ and $s'_0$ in
$\mbox{GeV}^2$.}\label{tab:s0}
\end{table}

\begin{figure}[tb]
$$\epsfxsize=0.60\textwidth\epsffile{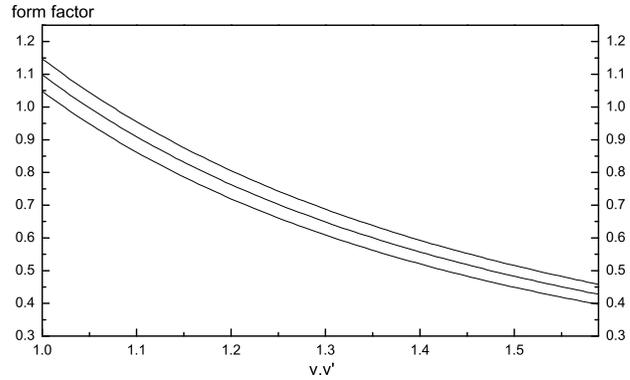}$$
\caption[]{Dependence of ${\cal F}_{B\to D}$ on the different sets
of parameters $m_b, f_B, s_0$. The three curves correspond to the
the parameter set $1{-}3$ from bottom to top. Here we use model
{I}{I} for the $D$-meson DA for simplicity.}\label{fig:f(mb)}
\end{figure}

\begin{figure}[p]
$$\epsfxsize=0.60\textwidth\epsffile{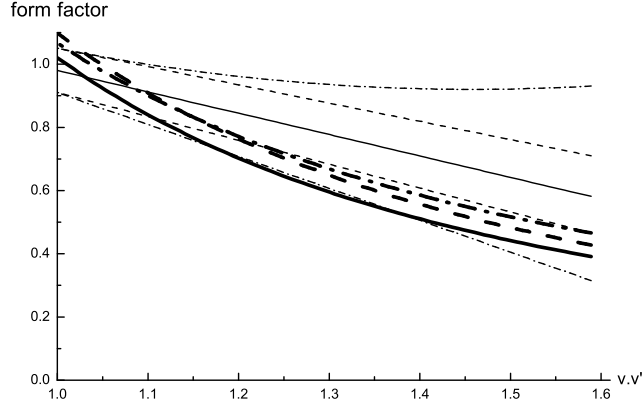}$$
\caption[]{${\cal F}_{B\to D}$ as a function of the velocity
transfer (with the parameters in the set 2). The thin lines
expresses the experiment fits results, the solid line represents the
central values, the dashed(dash-dotted) lines give the bounds from
the linear(quadratic) fits. The thick lines correspond to our
results, with the solid, dashed and dash-dotted lines for model
{I}{I}{I}, {I}{I} and I respectively.}\label{fig:F(y)}
\end{figure}

\begin{figure}[p]
$$\epsfxsize=0.60\textwidth\epsffile{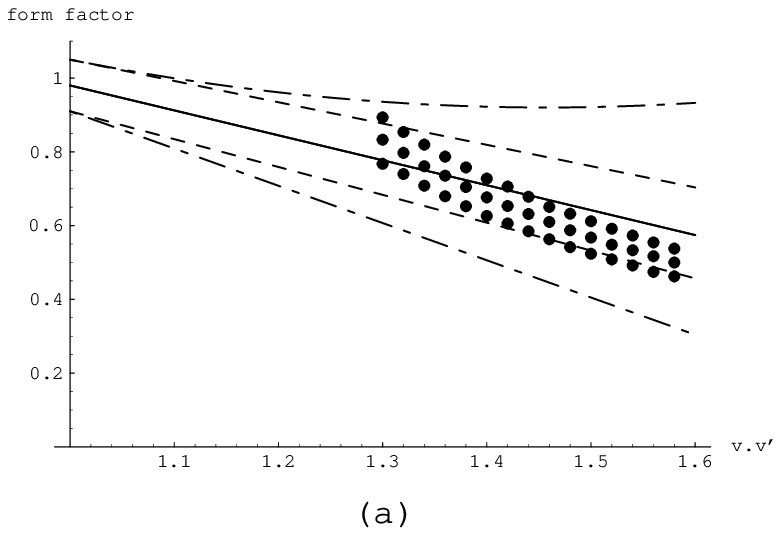}$$
\caption[]{pQCD results for ${\cal F}_{B\to D}(y)$ copied from
Ref.\cite{pQCD}. As in Fig.(\ref{fig:F(y)}), the solid line
represents the central values, the dashed(dash-dotted) lines give
the bounds from the linear(quadratic) fits. The circles corresponds
to pQCD results using model I for the $D$-meson DA with $C_D=0.5,
0.7, 0.9$ from bottom to top.}\label{fig:pQCD}
\end{figure}

With the parameters chosen, it is straightforward to calculate the
form factor ${\cal F}_{B\to D}(y)$ in the region $1.14<y<1.59$. The
results with different sets of parameters are plotted in
Fig.(\ref{fig:f(mb)}), where only model {I}{I} has been used for
simplicity. It is shown that the change of parameters can induce a
uncertainty of about $10-15\%$ if we let $m_b$ vary between
$4.75-4.85~\mbox{GeV}$. By fitting the data, the behavior of ${\cal
F}_{B\to D}(y)$ has been known using the parametrization
\begin{equation}
{\cal F}_{B\to D}(y)={\cal F}_{B\to
D}(1)[1-\hat{\rho}_D^2(y-1)+{\hat c}_D(y-1)^2+O((y-1)^3)],
\end{equation}
with
\begin{eqnarray}
\hat{\rho}_D^2&=&0.69\pm0.14,~~\hat c_D=0,\nonumber\\
\hat{\rho}_D^2&=&0.69^{+0.42}_{-0.15},~~\hat
c_D=0.00^{+0.59}_{-0.00},
\end{eqnarray}
corresponding to the linear and quadratic fits \cite{experiment},
respectively. With the three DA models, the resulting dependence of
${\cal F}_{B\to D}(y)$ on the velocity transfer $y$, along with that
extracted experimentally is illustrated in Fig.(\ref{fig:F(y)}). In
what follows, we denote the LCSR results for the form factor by
${\cal F}_{B\to D}^{LC}(y)$ and those extracted experimentally by
${\cal F}_{B\to D}^{exp}(y)$. For comparison, a figure-copy which
expresses the pQCD results in \cite{pQCD} is given in
Fig.(\ref{fig:pQCD}). In the region to which the LCSR method is
applicable, the central values of ${\cal F}_{B\to D}^{LC}(y)$ turn
out to be a bit smaller than the corresponding those of ${\cal
F}_{B\to D}^{exp}(y)$, using the DA models {I}{I} and {I}{I}{I} as
inputs; however, both of them are in accordance with each other
within the error. The situation with model I DA is about the same.
The central value of the form factor at the largest recoil is ${\cal
F}_{B\to D}^{exp}(1.59)=0.58$ versus ${\cal F}_{B\to
D}^{LC}(1.59)=0.40-0.50$, depending on the DA models. We note that
the behavior of ${\cal F}_{B\to D}^{LC}(y)$ is essentially unchanged
when the three different DA's are used. From the present
calculations, therefore it is too early to draw a conclusion which
DA model is more suitable to reflect the characteristics of QCD
dynamics inside the $D$ meson. When a comparison is made between the
pQCD and LCSR predictions, the consistent results can also be
observed at the larger recoil. Of course, the two approaches
describe the different dynamics in $ B\to D$ transitions. Whereas
the use of LCSR approach is to assume that the soft exchanges
dominate in the weak decay in question, applying pQCD method to do
calculation corresponds to the viewpoint that the hard exchanges do.

In fact, the kinematical region we give, which makes LCSR results
valid, is a conservative estimate. It is possible to extrapolate the
present LCSR calculation to the small recoil region. If it is true,
we find that in the whole kinematically accessible range $1.0\leq y
\leq 1.59$, the yielding LCSR estimates are compatible with the
data. For instance, at zero recoil it follows that ${\cal F}_{B\to
D}^{LC}(1)=1.02$ (using model {I}{I}{I}), which is in a good
agreement with the evaluation obtained using the heavy quark
symmetry: ${\cal F}_{B\to D}(1)=0.98\pm 0.07$. Nevertheless, we have
to emphasize that a full understanding of the dynamics involved in $
B\to D $ transition should be obtained by combining the three
different approaches~---~the lattice QCD calculations with the heavy
quark symmetry considered, LCSR results and pQCD predictions, which
are complementary to each other. The LCSR results with chiral
current correlator may act as a bridge connecting those of other
approaches.

\section{Summary}

~~~We have discussed the form factor for $B{\to}D$ transitions
${\cal F}_{B\to D}(y)$, using the improved QCD LCSR approach where
with the chiral current correlator chosen only the leading twist DA
of the $D$-meson is relevant at twist-3 accuracy. The resulting
LCSR's for ${\cal F}_{B\to D}(y)$ are available in the velocity
transfer region $1.14 < y < 1.59$. Calculation is done using three
different twist-2 DA models for D meson. It is shown the numerical
results are less sensitive to the choice of DA, and are of a central
value slight smaller than but within the error in a agreement with
those obtained by fitting the data on $ B\to D l\tilde{\nu}$. In the
larger recoil region $1.35 < y < 1.59$ where pQCD is applicable, the
results presented here are consistent with ones of pQCD. From the
practical calculations, we find that the present results might be
extrapolated to the smaller recoil region so that the $B{\to}D$
transitions are calculable in the whole kinematically accessible
range, using the improved LCSR approach.

Also, we argue that for understanding the form factor for $B{\to}D
l\tilde{\nu}$ in the whole kinematical range a combined use is
necessary of three different methods: the lattice QCD (with the
heavy quark symmetry considered), improved LCSR and pQCD approaches,
which are adequate to do calculation in different kinematical
regions and so could be complementary to each other. The LCSR
approach plays a bridge role in doing such calculation.

The present findings can be improved once the QCD radiative
correction to the LCSR is taken into account and a more reliable
twist-2 DA of $D$ meson becomes available. From the previous
discussion in \cite{cLCSR}, however, it is expected that the QCD
radiative correction can not change the present results too much.

\newpage

\begin{center}
{\bf ACKNOWLEDGEMENTS}
\end{center}

This work was supported in part by the Natural Science Foundation
of China (NSFC). We would like to thank Dr X. G. Wu for helpful discussions. \\

\end{document}